\documentclass[12pt]{article}

\usepackage{amsthm,amsmath,amssymb}
\usepackage{url}
\usepackage{srcltx} 
\usepackage{enumerate}
\usepackage{bm}
\usepackage[utf8]{inputenc}
\usepackage{mathrsfs}
\usepackage{xspace}
\usepackage{framed}
\usepackage{enumerate}
\usepackage{xcolor}
\usepackage{xspace}
\usepackage{graphicx}
\usepackage[colorlinks=true,citecolor=black,linkcolor=black,urlcolor=blue]{hyperref}
\usepackage{authblk}

\DeclareMathOperator{\pexp}{pexp}

\def\paren#1{\left( #1 \right)}
\def\acc#1{\left\{ #1 \right\}}

\renewcommand{\le}{\leqslant}
\renewcommand{\ge}{\geqslant}

\newenvironment{proofof}[1]{\par \noindent \textit{Proof of #1.}
}{\hfill$\Box$\medskip} 

\theoremstyle{plain}
\newtheorem{theorem}{Theorem}

\newtheorem{proposition}[theorem]{Proposition}

\theoremstyle{definition}

\theoremstyle{remark}



\title{Repetition avoidance in products of factors}

\author{
Pamela Fleischmann\thanks{Kiel University, Germany. \texttt{fpa@informatik.uni-kiel.de}} \quad
Pascal Ochem\thanks{LIRMM, Universit\'e de Montpellier, CNRS, Montpellier, France. \texttt{ochem@lirmm.fr}} \quad
Kamellia Reshadi\thanks{Kiel University, Germany. \texttt{kre@informatik.uni-kiel.de}} \quad}

\date{}

\begin{document}

\maketitle

\begin{abstract}
We consider a variation on a classical avoidance problem from combinatorics on words
that has been introduced by Mousavi and Shallit at DLT 2013.
Let $\texttt{pexp}_i(w)$ be the supremum of the exponent over the products (concatenation) of $i$ factors of the word $w$.
The repetition threshold $\textsc{RT}_i(k)$ is then the infimum of $\texttt{pexp}_i(w)$ over all words $w\in\Sigma^\omega_k$.
Mousavi and Shallit obtained that $\textsc{RT}_i(2)=2i$ and $\textsc{RT}_2(3)=\tfrac{13}4$.
We show that $\textsc{RT}_i(3)=\tfrac{3i}2+\tfrac14$ if $i$ is even and $\textsc{RT}_i(3)=\tfrac{3i}2+\tfrac16$ if $i$ is odd and $i\ge3$.

\end{abstract}

\textbf{Acknowledgements}: This work was partially supported by the ANR project CoCoGro (ANR-16-CE40-0005).

\section{Introduction}\label{sec:intro}

A \emph{repetition} in a word $w$ is a pair of words $p$ and $e$ such that
$pe$ is a factor of $w$, $p$ is non-empty, and $e$ is a prefix of $pe$.
If $pe$ is a repetition, then its \emph{period} is $|p|$ and its \emph{exponent} is $\tfrac{|pe|}{|p|}$.
A word is \emph{$\alpha^+$-free} (resp. \textit{$\alpha$-free}) if it 
contains no repetition with exponent $\beta$ such that $\beta>\alpha$ (resp. $\beta\ge\alpha$).

\medskip

Given $k\ge 2$, Dejean~\cite{Dejean:1972} defined the repetition threshold
$\textsc{RT}(k)$ for $k$ letters as the smallest $\alpha$ such that there
exists an infinite $\alpha^+$-free word over a $k$-letter alphabet $\Sigma_k=\acc{0,1,\ldots,k-1}$.
Dejean initiated the study of $\textsc{RT}(k)$ in 1972 for $k=2$ and $k=3$.
Her work was followed by a series of papers which determine the exact
value of $\textsc{RT}(k)$ for any $k\ge 2$.

\begin{itemize}
\item $\textsc{RT}(2) = 2$~\cite{Dejean:1972};
\item $\textsc{RT}(3) = \tfrac{7}{4}$~\cite{Dejean:1972};
\item $\textsc{RT}(4) = \tfrac{7}{5}$~\cite{Pan84};
\item $\textsc{RT}(k) = \tfrac{k}{k-1}$, for $k\ge 5$~\cite{Car07,Oll92,Rao11}.
\end{itemize}
Mousavi and Shallit~\cite{MS13} have considered two notions related to the repetition threshold.

The first notion considers repetitions in conjugates of factors of the infinite word.
A word is circularly $r^+$-free if it does not contain a factor $pxs$ such that $sp$ is a repetition of exponent strictly greater than $r$.
The smallest real number $r$ such that $w$ is circularly $r^+$-free is denoted by $\texttt{cexp}(w)$.
Let $\textsc{RTC}(k)$ be the minimum of $\texttt{cexp}(w)$ over every $w\in\Sigma_k^\omega$.

The second notion considers repetitions in concatenations of a fixed number of factors of the infinite word.
Let $\texttt{pexp}_i(w)$ be the smallest real number $r$ such that every product of $i$ factors of $w$ is $r^+$-free.
Let $\textsc{RT}_i(k)$ be the minimum of $\texttt{pexp}_i(w)$ over every $w\in\Sigma_k^\omega$.
Notice that $\textsc{RT}_i(k)$ generalizes the classical notion of repetition threshold which corresponds to the case $i=1$, that is,
$\textsc{RT}_1(k)=\textsc{RT}(k)$ for every $k\ge2$.

For the case $i=2$, Mousavi and Shallit obtained the following.

\begin{proposition}~\cite{MS13}\label{prop:ms}
If $w$ is a recurrent infinite word, then $pexp_2(w)=cexp(w)$.
\end{proposition}

Notice that the language of circularly $r^+$-free words in $\Sigma_k^*$ is a factorial language.
As it is well-known~\cite{Fogg}, if a factorial language is infinite, then it contains a uniformly recurrent word.
Thus, $\textsc{RTC}(k)$ can be equivalently defined as the minimum of $\texttt{cexp}(w)$ over every uniformly recurrent word $w\in\Sigma_k^\omega$.
Then Proposition~\ref{prop:ms} implies the following result.

\begin{proposition}\label{prop:eq}
For every $k\ge2$, $\textsc{RT}_2(k)=\textsc{RTC}(k)$.
\end{proposition}

Mousavi and Shallit~\cite{MS13} have considered the binary alphabet and obtained that $\textsc{RT}_i(2)=2i$ for every $i\ge1$.
Our main result considers the ternary alphabet and gives the value of $\textsc{RT}_i(3)$ for every $i\ge1$.
This extends the result of Dejean~\cite{Dejean:1972} that $\textsc{RT}_1(3)=\tfrac74$
and the result of Mousavi and Shallit~\cite{MS13} that $\textsc{RT}_2(3)=\tfrac{13}4$.

\begin{theorem}\label{thm:main}{\ }
\begin{itemize}
 \item $\textsc{RT}_i(3)=\tfrac{3i}2+\tfrac14$ if $i=1$ or $i$ is even.
 \item $\textsc{RT}_i(3)=\tfrac{3i}2+\tfrac16$ if $i$ is odd and $i\ge3$.
\end{itemize} 
\end{theorem}

\section{Proofs}\label{sec:proofs}

To obtain the two equalities of Theorem~\ref{thm:main}, we show the two lower bounds and then the two upper bounds.\\

\begin{proofof}{ $\textsc{RT}_i(3)\ge\tfrac{3i}2+\tfrac14$ for every even $i$}\\
Mousavi and Shallit~\cite{MS13} have proved that $\textsc{RT}_2(3)=\tfrac{13}4$, which settles the case $i=2$.
We have double checked their computation of the lower bound $\textsc{RT}_2(3)\ge\tfrac{13}4$.
Suppose that $i$ is a fixed even integer and that $w_3$ is an infinite ternary word.
The lower bound for $i=2$ implies that there exists two factors $u$ and $v$ such that $uv=t^e$ with $e\ge\tfrac{13}4$.
Thus, the prefix $t^3$ of $uv$ is also a product of two factors of $w_3$.
So we can form the $i$-terms product $(t^3)^{i/2-1}uv$ which is a repetition of the form $t^x$
with exponent $x=3\paren{\tfrac i2-1}+e\ge3\paren{\tfrac i2-1}+\tfrac{13}4=\tfrac{3i}2+\tfrac14$.
This is the desired lower bound.
\end{proofof}

\medskip

\begin{proofof}{ $\textsc{RT}_i(3)\ge\tfrac{3i}2+\tfrac16$ for every odd $i\ge3$}\\
Suppose that $i\ge3$ is a fixed odd integer, that is, $i=2j+1$.
Suppose that $w_3$ is a recurrent ternary word such that the product of $i$ factors of $w_3$
is never a repetition of exponent at least $\tfrac{3i}2+\tfrac16=3j+\tfrac53$.
First, $w_3$ is square-free since otherwise there would exist an $i$-terms product of exponent $2i$.
Also, $w_3$ does not contain two factors $u$ and $v$ with the following properties:
\begin{itemize}
 \item $uv=t^3$,
 \item $u=t^e$ with $e\ge\tfrac53$.
\end{itemize}
Indeed, this would produce the $i$-terms product $(uv)^ju$ which is a repetition of the form $t^x$ with exponent
$x=3j+e\ge3j+\tfrac53$.

So if $a$, $b$, and $c$ are distinct letters, then $w_3$ does not contain both $u=abcab$ and $v=cabc$
and $w_3$ does not contain both $u=abcbabc$ and $v=babcb$.
A computer check shows that no infinite ternary square-free word satisfies this property.
This proves the desired lower bound.
\end{proofof}

\medskip

\begin{proofof}{ $\textsc{RT}_i(3)\le\tfrac{3i}2+\tfrac14$ for every even $i$}\\
Let $i$ be any even integer at least $2$. 
To prove this upper bound, it is sufficient to construct a ternary word $w$ satisfying $\texttt{pexp}_i(w)\le\tfrac{3i}2+\tfrac14$.
The ternary morphic word used in~\cite{MS13} to obtain $\textsc{RT}_2(3)\le\tfrac{13}4$ seems to satisfy the property.
However, it is easier for us to consider another construction.
Let us show that the image of every $7/5^+$-free word over $\Sigma_4$ by the following $45$-uniform morphism satisfies
$\texttt{pexp}_i\le\tfrac{3i}2+\tfrac14$.

$$
\begin{array}{ll}
\texttt{0}\mapsto&\texttt{010201210212021012102010212012101202101210212}\\
\texttt{1}\mapsto&\texttt{010201210212012101202101210201021202101210212}\\
\texttt{2}\mapsto&\texttt{010201210120212012102120210121021201210120212}\\
\texttt{3}\mapsto&\texttt{010201210120210121021201210120212012102010212}\\
\end{array}
$$

Recall that a word is $(\beta^+,n)$-free if it does not contain a repetition
with period at least $n$ and exponent strictly greater than $\beta$.
First, we check that such ternary images are $\paren{\tfrac{202}{135}^+,36}$-free using the method in~\cite{Ochem2004}.
By Lemma~2.1 in~\cite{Ochem2004}, it is sufficient to check this freeness property for the image of every $7/5^+$-free word over $\Sigma_4$ of length smaller than
$\frac{2\times\tfrac{202}{135}}{\tfrac{202}{135}-\tfrac75}<32$.
Since $\tfrac{202}{135} < \tfrac32$, the period of every repetition formed from $i$ pieces and with exponent at least $\tfrac{3i}2$ 
must be at most $35$.
Then we check exhaustively by computer that the ternary images do not contain two factors $u$ and $v$ such that
\begin{itemize}
 \item $uv=t^e$,
 \item $e>3$,
 \item $9\le |t|\le 35$.
\end{itemize}
Thus, the period of every repetition formed from $i$ pieces and with exponent strictly greater than $\tfrac{3i}2$ must be at most $8$.
So we only need to check that $\texttt{pexp}_i\le\tfrac{3i}2+\tfrac14$ for $i$-terms products that are repetitions of period at most $8$.

Now the period is bounded, but $i$ can still be arbitrarily large, a priori.
For every factor $t$ of length at most $8$, we define $\texttt{pexp}_{i,t}$ as the length of a largest factor of $t^\omega$ that is a $i$-terms product, divided by $|t|$.
We actually consider conjugacy classes, since if $t'$ is a conjugate of $t$, then $\texttt{pexp}_{i,t'}=\texttt{pexp}_{i,t}$.
Let $t$ be such a factor. If, for some even $j$, we have $\texttt{pexp}_{j+2,t}=\texttt{pexp}_{j,t}+3$, then it means that by appending a $2$-terms product
to a $j$-terms product that corresponds to a maximum factor of $t^\omega$, that can only add a cube of period $|t|$.
This implies that for every $k$, $\texttt{pexp}_{j+2k,t}=\texttt{pexp}_{j,t}+3k$.

We have checked by computer that for every conjugacy class of words $t$ of length at most $8$,
there exists a (small) even $j$ such that $\texttt{pexp}_{j+2,t}=\texttt{pexp}_{j,t}+3$.
Thus we have $\texttt{pexp}_i\le\tfrac{3i}2+\tfrac14$ in all cases.
\end{proofof}

\medskip

\begin{proofof}{ $\textsc{RT}_i(3)\le\tfrac{3i}2+\tfrac16$ for every odd $i\ge3$}\\
Let us show that the image of every $7/5^+$-free word over $\Sigma_4$ by the following $514$-uniform
morphism satisfies $\texttt{pexp}_i\le\tfrac{3i}2+\tfrac16$ for every odd $i\ge3$.
{\small
$$
\begin{array}{ll}
\texttt{0}\mapsto&\texttt{01020120210120102120210201210120102012021020121021201020121012}\\
&\texttt{02102012102120210120102012102120102012021020121012010212021020}\\
&\texttt{12102120102012021012010212021020121021202101201020121021201020}\\
&\texttt{12101202102012102120210120102120210201210120102012021020121012}\\
&\texttt{01021202102012102120102012101202102012102120102012021012010212}\\
&\texttt{02102012101201020120210201210212021012010201210120210201210212}\\
&\texttt{01020120210201210120102120210201210212010201210120210201210212}\\
&\texttt{02101201021202102012101201020120210201210120102120210201210212}\\
&\texttt{021012010201210212}\\
\end{array}
$$

$$
\begin{array}{ll}
\texttt{1}\mapsto&\texttt{01020120210120102120210201210120102012021020121021201020121012}\\
&\texttt{02102012102120102012021020121012010212021020121021201020120210}\\
&\texttt{12010212021020121021202101201020121021201020121012021020121021}\\
&\texttt{20210120102120210201210120102012021020121021201020121012021020}\\
&\texttt{12101201021202102012102120210120102012102120102012021012010212}\\
&\texttt{02102012101201020120210201210120102120210201210212010201210120}\\
&\texttt{21020121021202101201021202102012101201020120210201210212021012}\\
&\texttt{01020121012021020121021201020120210201210120102120210201210212}\\
&\texttt{021012010201210212}\\
\end{array}
$$

$$
\begin{array}{ll}
\texttt{2}\mapsto&\texttt{01020120210120102120210201210120102012021020121021201020121012}\\
&\texttt{02102012101201021202102012102120102012021012010212021020121021}\\
&\texttt{20210120102012102120102012101202102012102120210120102120210201}\\
&\texttt{21012010201202102012101201021202102012102120102012101202102012}\\
&\texttt{10212010201202101201021202102012102120210120102012102120102012}\\
&\texttt{02102012101201021202102012102120102012021012010212021020121012}\\
&\texttt{01020120210201210212021012010201210212010201210120210201210212}\\
&\texttt{02101201021202102012101201020120210201210120102120210201210212}\\
&\texttt{021012010201210212}\\
\end{array}
$$

$$
\begin{array}{ll}
\texttt{3}\mapsto&\texttt{01020120210120102120210201210120102012021020121021201020121012}\\
&\texttt{02102012101201021202102012102120102012021012010212021020121021}\\
&\texttt{20210120102012101202102012102120102012021020121012010212021020}\\
&\texttt{12102120102012101202102012102120210120102012102120102012021012}\\
&\texttt{01021202102012101201020120210201210212010201210120210201210212}\\
&\texttt{01020120210201210120102120210201210212021012010201210212010201}\\
&\texttt{20210120102120210201210212010201210120210201210212021012010212}\\
&\texttt{02102012101201020120210201210120102120210201210212021012010201}\\
&\texttt{210120210201210212}\\
\end{array}
$$
}

First, we check that such ternary images are $\paren{\tfrac32^+,45}$-free using the method in~\cite{Ochem2004}.
By Lemma 2.1 in~\cite{Ochem2004}, it is sufficient to check this freeness property for the image of every $7/5^+$-free word over $\Sigma_4$ of length smaller than
$\frac{2\times\tfrac32}{\tfrac32-\tfrac75}=30$.
Thus, the period of every repetition formed from $i$ pieces and with exponent strictly greater than $\tfrac{3i}2$ must be at most $44$.
Using the same argument as in the previous proof, we have checked by computer that for every conjugacy class of words $t$ of length at most $44$,
there exists a (small) odd $j$ such that $\texttt{pexp}_{j+2,t}=\texttt{pexp}_{j,t}+3$.
Thus we have $\texttt{pexp}_i\le\tfrac{3i}2+\tfrac16$ in all cases.
\end{proofof}

Let us describe how the morphisms above were found.
For increasing $k$, we try to find a $k$-uniform morphism $m$ by looking for a ternary square-free word $w$ of length $4k$
(with the suitable $\pexp_i(w)$ properties) that corresponds to $m(\texttt{0123})$.
We use the following optimizations to speed up the backtracking.
\begin{itemize}
 \item We force $m(\texttt{0}) > m(\texttt{1}) > m(\texttt{2}) > m(\texttt{3})$ with respect to the lexicographic order.
 \item we use early tests: 
 \begin{itemize}
  \item if we have a candidate for $m(\texttt{01})$, then we also test $m(\texttt{10})$; 
  \item if we have a candidate for $m(\texttt{012})$, then we also test every word $m(abca)$ such that $\acc{a,b,c}=\acc{\texttt{0},\texttt{1},\texttt{2}}$
 \end{itemize}
\end{itemize}
The general idea of the method is that large occurrences of the forbidden structures are ruled out thanks to an argument about the 
exponent of the repetitions induced by these structures.
Then the small occurrences are ruled out by an exhaustive inspection of the factors of the word of some finite length.

\section{Concluding remarks}\label{sec:con}
The next step would be to consider the $4$-letter alphabet.
Obviously, $\textsc{RT}_{i+1}(k)\ge\textsc{RT}_i(k)+1$ for every $i\ge1$ and $k\ge2$.
Mousavi and Shallit~\cite{MS13} verified that $\textsc{RT}_2(4)\ge\tfrac52$,
so that $\textsc{RT}_i(4)\ge i+\tfrac12$ for every $i\ge2$.
We conjecture that this is best possible, i.e., that $\textsc{RT}_i(4)=i+\tfrac12$ for every $i\ge2$.
However, a proof of an upper bound of the form $\textsc{RT}_i(4)\le i+c$
cannot be similar to the proof of the upper bounds of Theorem~\ref{thm:main}.
The multiplicative factor of $i$, which drops from $\tfrac32$ when $k=3$ to $1$ when $k=4$,
forbids that the constructed word is the morphic image of any (unspecified) Dejean word over a given alphabet.

Proving some of the conjectured values of $\textsc{RT}_i$ would
lead to stronger versions of the classical repetition threshold:
every witness of $\textsc{RT}_i(k)=\textsc{RT}(k)+i-1$ is a Dejean word with severe 
restrictions on the types of repetitions that are allowed to appear.

\end{document}